\documentclass[12pt]{article}
\usepackage{amsmath,amssymb,amsthm,amsxtra,overpic,bbm,bm,epsfig,ulem}
\usepackage{color}
\usepackage{cite}

\usepackage{hyperref}
\usepackage{url}

\def\thefootnote{\fnsymbol{footnote}}

\begin{document}

\vspace{0.2cm}

\begin{center}
{\Large\bf The formal seesaw mechanism of Majorana neutrinos \\
with unbroken gauge symmetry}
\end{center}

\vspace{0.2cm}

\begin{center}
{\bf Zhi-zhong Xing$^{1,2}$}
\footnote{E-mail: xingzz@ihep.ac.cn}
\\
{\small $^{1}$Institute of High Energy Physics and School of Physical Sciences, \\
University of Chinese Academy of Sciences, Beijing 100049, China \\
$^{2}$Center of High Energy Physics, Peking University, Beijing 100871, China}
\end{center}

\vspace{2cm}
\begin{abstract}
We reformulate the canonical seesaw mechanism in the case that the
electroweak gauge symmetry is unbroken, and show that it can {\it formally}
work and allow us to derive an exact seesaw formula for the light and
heavy Majorana neutrinos. We elucidate the reason why there is a mismatch
between the mass eigenstates of heavy Majorana neutrinos associated with
thermal leptogenesis and those associated with the seesaw framework, and
establish the exact and explicit relations between the {\it original} and
{\it derivational} seesaw parameters by using an Euler-like parametrization
of the $6\times 6$ active-sterile flavor mixing matrix.
\end{abstract}

\newpage

\def\thefootnote{\arabic{footnote}}
\setcounter{footnote}{0}

\setcounter{equation}{0}
\section{Motivation}

Among all the proposed mechanisms toward deeply understanding the true origin
of tiny masses of the three known neutrinos $\nu^{}_i$ (for $i = 1, 2, 3$),
whose flavor eigenstates are commonly denoted as $\nu^{}_\alpha$ (for
$\alpha = e, \mu, \tau$), the canonical seesaw mechanism~\cite{Minkowski:1977sc,
Yanagida:1979as,GellMann:1980vs,Glashow:1979nm,Mohapatra:1979ia} stands out as
being most economical and most natural. The simplicity of this mechanism lies
in two aspects: (a) it just takes into account the right-handed neutrino fields
$N^{}_{\alpha \rm R}$, the chiral counterparts of the left-handed neutrino fields
$\nu^{}_{\alpha \rm L}$ (for $\alpha = e, \mu, \tau$), which were originally
ignored from the particle content of the standard model (SM)~\cite{Weinberg:1967tq};
(b) it simply allows for lepton number violation or the Majorana nature of
massive neutrinos~\cite{Majorana:1937vz}, which is completely
harmless to the theoretical framework of the SM itself. The naturalness of this
mechanism is reflected in its attributing the small masses of $\nu^{}_i$
to the existence of three heavy Majorana neutrinos $N^{}_i$ (for $i = 1, 2, 3$),
whose masses are expected to be far above the fulcrum of the seesaw --- presumably
the electroweak symmetry breaking scale of the SM characterized by the vacuum
expectation value of the Higgs field. On the other hand,
the seesaw mechanism offers a big bonus to cosmology: the CP-violating and
out-of-equilibrium decays of heavy Majorana neutrinos may give rise to a net
lepton-antilepton asymmetry in the early Universe, and such a {\it leptogenesis}
mechanism~\cite{Fukugita:1986hr} can finally lead to {\it baryogenesis} as a
natural interpretation of the observed baryon-antibaryon asymmetry in today's
Universe~\cite{ParticleDataGroup:2022pth}. In this sense the seesaw
mechanism is the very {\it stone} that can kill two fundamental {\it birds}
in particle physics and cosmology.

Note that the seesaw mechanism is expected to take effect at a superhigh energy
scale $\Lambda$ which is essentially of the order of the heavy Majorana neutrino
masses. But the $\rm SU(2)^{}_{\rm L} \times U(1)^{}_{\rm Y}$ electroweak
gauge symmetry has been unbroken until the Higgs field develops a nonzero
vacuum expectation value $v$ of ${\cal O}\big(10^2\big)~{\rm GeV}$. In this
situation the three active neutrinos are actually impossible to acquire their
{\it true} masses of ${\cal O}(v^2/\Lambda)$ at the seesaw scale $\Lambda$ due to
the absence of a {\it real} fulcrum of the seesaw. On the other hand, thermal
leptogenesis can be realized via the lepton-number-violating decays of heavy
Majorana neutrinos into the leptonic and Higgs doublets at $\Lambda$. So we
are well motivated to ask a conceptually important question: how can the seesaw
mechanism {\it formally} survive with the unbroken electroweak gauge symmetry
and work together with the leptogenesis mechanism? If the answer to this question
is affirmative, we wonder whether the mass eigenstates of heavy Majorana neutrinos
associated with thermal leptogenesis are exactly the same as those associated
with the seesaw mechanism itself
%%%%%%%%%%%%%%%%%%%%%%%%%%%%%%%%%%%%%%%%%%%%%%%%%%%%%%%%%%%%%%%%%%%%%%%%%%%%%%%
\footnote{A mismatch of this kind has been observed and discussed in the seesaw
framework {\it after} spontaneous electroweak symmetry breaking and in an
{\it approximate} way (see, e.g., Refs.~\cite{Drewes:2013gca,Canetti:2012kh,
Drewes:2019byd,Chrzaszcz:2019inj,Klaric:2021cpi}). Here we shall take a new look
at it {\it before} electroweak symmetry breaking and in an {\it exact} way at
the tree level.}.
%%%%%%%%%%%%%%%%%%%%%%%%%%%%%%%%%%%%%%%%%%%%%%%%%%%%%%%%%%%%%%%%%%%%%%%%%%%%%%%
In case that there exists a mismatch between these two sets of mass bases, then
the question becomes how small this mismatch is likely to be.

To answer the above questions and clarify some conceptual ambiguities that have
never been taken seriously, we are going to study how to make the seesaw mechanism
formally work before spontaneous electroweak symmetry breaking. We show that an
exact seesaw relation between the light and heavy Majorana neutrinos can be
established far above the electroweak scale, and it becomes the realistic seesaw
relation after the Higgs field develops its vacuum expectation value.
In this way it is straightforward to elucidate the reason why there is a
mismatch between the mass eigenstates of heavy Majorana neutrinos associated
with thermal leptogenesis and those associated with the seesaw mechanism. With
the help of a full Euler-like parametrization of the flavor structure in the
seesaw framework, we illuminate such a mismatch in a more specific way.
The exact and explicit relations between the {\it original} and {\it derivational}
parameters of massive Majorana neutrinos are obtained as a by-product, and
they are expected to be useful in determining or constraining some of the
original seesaw parameters from the low-energy neutrino experiments.

\setcounter{equation}{0}
\section{A formal seesaw mechanism?}

\subsection{The leptonic Yukawa interactions}

Let us begin with the gauge-invariant leptonic Yukawa interactions and the
$\rm SU(2)^{}_{\rm L}$-singlet Majorana neutrino mass term of the canonical
seesaw mechanism at $\Lambda$
%%%%%%%%%%%%%%%%%%%%%%%%%%%%%%%%%%%%%%%%%%%%%%%%%%%%%%%%%%%%%%%%%%%%%%%%%%%%
\footnote{Throughout this paper, our discussions are subject to the minimal
extension of the SM with three right-handed neutrino fields and lepton number
violation at {\it zero} temperature, so as to make our key point clear and
avoid possible complications (e.g., thermal corrections to the masses of
heavy Majorana neutrinos~\cite{Klaric:2021cpi}).}
%%%%%%%%%%%%%%%%%%%%%%%%%%%%%%%%%%%%%%%%%%%%%%%%%%%%%%%%%%%%%%%%%%%%%%%%%%%%
\begin{eqnarray}
-{\cal L}^{}_{\Lambda} = \overline{\ell^{}_{\rm L}} \hspace{0.05cm} Y^{}_l
H \hspace{0.02cm} l^{}_{\rm R} + \overline{\ell^{}_{\rm L}} \hspace{0.05cm}
Y^{}_\nu \widetilde{H} N^{}_{\rm R}
+ \frac{1}{2} \hspace{0.05cm}
\overline{(N^{}_{\rm R})^c} \hspace{0.05cm} M^{}_{\rm R} N^{}_{\rm R}
+ {\rm h.c.} \; ,
\label{1}
%     (1)
\end{eqnarray}
where $\ell^{}_{\rm L} = \big(\begin{matrix} \nu^{}_{\rm L} & l^{}_{\rm L}
\end{matrix}\big)^T$ denotes the leptonic $\rm SU(2)^{}_{\rm L}$ doublet of
the SM with $\nu^{}_{\rm L} = \big(\begin{matrix} \nu^{}_{e \rm L} &
\nu^{}_{\mu \rm L} & \nu^{}_{\tau \rm L}\end{matrix}\big)^T$ and
$l^{}_{\rm L} = \big(\begin{matrix} l^{}_{e \rm L} & l^{}_{\mu \rm L} &
l^{}_{\tau \rm L}\end{matrix}\big)^T$ standing respectively for the column
vectors of the left-handed neutrino and charged lepton fields,
$\widetilde{H} \equiv {\rm i} \sigma^{}_2 H^*$ with
$H = \big(\begin{matrix} \phi^+ & \phi^0\end{matrix}\big)^T$ being the Higgs
doublet of the SM and $\sigma^{}_2$ being the second Pauli matrix,
$l^{}_{\rm R} = \big(\begin{matrix} l^{}_{e \rm R} & l^{}_{\mu \rm R} &
l^{}_{\tau \rm R}\end{matrix}\big)^T$ and $N^{}_{\rm R} = \big(\begin{matrix}
N^{}_{e \rm R} & N^{}_{\mu \rm R} & N^{}_{\tau \rm R}\end{matrix}\big)^T$
stand respectively for the column vectors of the right-handed charged lepton
and neutrino fields which are the $\rm SU(2)^{}_{\rm L}$
singlets, $(N^{}_{\rm R})^c \equiv {\cal C} \overline{N^{}_{\rm R}}^T$
with $\cal C$ being the charge-conjugation matrix and satisfying
${\cal C}^{-1} = {\cal C}^\dagger = {\cal C}^T = - {\cal C}$, $Y^{}_l$ and
$Y^{}_\nu$ represent the respective Yukawa coupling matrices of charged leptons
and neutrinos, and $M^{}_{\rm R}$ is the symmetric right-handed neutrino mass
matrix. In Eq.~(\ref{1}) the hypercharges of $\ell^{}_{\rm L}$, $l^{}_{\rm R}$,
$N^{}_{\rm R}$, $H$ and $\widetilde{H}$ are $-1/2$, $-1$, $0$, $+1/2$ and
$-1/2$, respectively.
Since $\overline{\nu^{}_{\rm L}} \hspace{0.05cm} Y^{}_\nu N^{}_{\rm R}$
is a Lorentz scalar and can be transformed into
\begin{eqnarray}
\overline{\nu^{}_{\rm L}} \hspace{0.05cm} Y^{}_\nu N^{}_{\rm R} = \left[
\overline{\nu^{}_{\rm L}} \hspace{0.05cm} Y^{}_\nu N^{}_{\rm R}\right]^T =
\overline{(N^{}_{\rm R})^c} \hspace{0.05cm} Y^T_\nu (\nu^{}_{\rm L})^c \; ,
\label{2}
%     (2)
\end{eqnarray}
where $(\nu^{}_{\rm L})^c \equiv {\cal C} \overline{\nu^{}_{\rm L}}^T$ is
the charge-conjugated counterpart of $\nu^{}_{\rm L}$, one may easily
rewrite Eq.~(\ref{1}) as
\begin{eqnarray}
-{\cal L}^{}_{\Lambda} \hspace{-0.2cm} & = & \hspace{-0.2cm}
\overline{l^{}_{\rm L}} \hspace{0.05cm} Y^{}_l \hspace{0.03cm}
l^{}_{\rm R} \phi^0 + \frac{1}{2} \hspace{0.05cm}
\overline{\big[\begin{matrix} \nu^{}_{\rm L} & (N^{}_{\rm R})^c\end{matrix}
\big]} \left(\begin{matrix} {\bf 0} & Y^{}_\nu \phi^{0*} \cr
Y^T_\nu \phi^{0*} & M^{}_{\rm R} \end{matrix}\right)
\left[\begin{matrix} (\nu^{}_{\rm L})^c \cr N^{}_{\rm R} \end{matrix}\right]
\hspace{0.5cm}
\nonumber \\
&& \hspace{-0.25cm} + \hspace{0.1cm}
\overline{\nu^{}_{\rm L}} \hspace{0.05cm} Y^{}_l \hspace{0.03cm} l^{}_{\rm R}
\phi^+ - \overline{l^{}_{\rm L}} \hspace{0.05cm} Y^{}_\nu N^{}_{\rm R} \phi^-
+ {\rm h.c.} \; .
\label{3}
%     (3)
\end{eqnarray}
This expression is highly nontrivial in the sense that it
clearly shows a direct correlation between the left- and right-handed neutrino
fields via their Yukawa couplings to the neutral component of the Higgs doublet
even though the $\rm SU(2)^{}_{\rm L} \times U(1)^{}_{\rm Y}$ gauge symmetry is
perfect at the seesaw scale $\Lambda$. In this situation the $3\times 3$
Yukawa coupling matrix $Y^{}_\nu$ can be regarded as a ``virtual" fulcrum of
the seesaw before spontaneous electroweak symmetry breaking.

Note that both the scalar field $\phi^0$ and its charge-conjugated counterpart
$\phi^{0 *}$ have the mass dimension and act like two complex numbers in
Eq.~(\ref{3}). But of course they possess the respective hypercharges $+1/2$
and $-1/2$ as $\phi^{\pm}$ do. After spontaneous symmetry breaking
$\phi^0$ and $\phi^{0 *}$ will acquire the same vacuum expectation value
$\langle \phi^0\rangle = \langle \phi^{0 *}\rangle = v/\sqrt{2}$ with
$v \simeq 246~{\rm GeV}$, together with $\langle \phi^-\rangle =
\langle \phi^+\rangle =0$, as in the SM. Then the formal seesaw will acquire
a real fulcrum which allows one to naturally attribute the smallness of
three active Majorana neutrino masses to the existence of three heavy
Majorana neutrinos, as can be seen later on.

\subsection{The leptogenesis-associated basis}

Now that all the SM particles are exactly massless in the early Universe when
the temperature is far above the electroweak scale, a realization of
thermal leptogenesis at $\Lambda \gg v$ only needs to calculate the
lepton-number-violating decays of heavy Majorana neutrinos into the
leptonic doublet and the Higgs doublet at the one-loop level by simply
starting from Eq.~(\ref{1}) instead of Eq.~(\ref{3}) (see, e.g.,
Refs.~\cite{Fukugita:1986hr,Luty:1992un,Covi:1996wh,Plumacher:1996kc,
Pilaftsis:1997jf}). In this case the column vector of the mass eigenstates of
three heavy Majorana neutrinos, denoted as
${\cal N}^\prime = \big(\begin{matrix} {\cal N}^{}_1 &
{\cal N}^{}_2 & {\cal N}^{}_3 \end{matrix}\big)^T$, can easily be obtained by
making the Autonne-Takagi transformation~\cite{Autonne1915,Takagi1924} as
follows:
\begin{eqnarray}
U^{\prime \dagger}_0 M^{}_{\rm R} U^{\prime *}_0 = {\cal D}^{}_{\cal N} \; , \quad
{\cal N}^\prime_{\rm R} = U^{\prime T}_0 N^{}_{\rm R} \; ,
\label{4}
%     (4)
\end{eqnarray}
where $U^\prime_0$ is a unitary matrix, and ${\cal D}^{}_{\cal N} \equiv
{\rm Diag}\big\{{\cal M}^{}_1 , {\cal M}^{}_2 , {\cal M}^{}_3 \big\}$ with
${\cal M}^{}_i$ being the masses of ${\cal N}^{}_i$ (for $i = 1, 2, 3$).
As a result, the Lagrangian ${\cal L}^{}_\Lambda$ in Eq.~(\ref{1}) becomes
\begin{eqnarray}
-{\cal L}^{}_{\Lambda} = \overline{\ell^{}_{\rm L}} \hspace{0.05cm} Y^{}_l
H \hspace{0.02cm} l^{}_{\rm R} + \overline{\ell^{}_{\rm L}} \hspace{0.05cm}
{\cal Y}^{}_\nu \widetilde{H} {\cal N}^{\prime}_{\rm R} + \frac{1}{2}
\hspace{0.05cm} \overline{({\cal N}^{\prime}_{\rm R})^c} \hspace{0.05cm}
{\cal D}^{}_{\cal N} {\cal N}^{\prime}_{\rm R}
+ {\rm h.c.} \; ,
\label{5}
%     (5)
\end{eqnarray}
where ${\cal Y}^{}_\nu \equiv Y^{}_\nu U^{\prime *}_0$ is defined for the sake
of simplicity. The rates of ${\cal N}^{}_i$ decaying into $\ell^{}_{\rm L}$ and
$H$ or their CP-conjugated states
are therefore determined by ${\cal M}^{}_i$ and ${\cal Y}^{}_\nu$, so are the
corresponding CP-violating asymmetries associated closely with thermal
leptogenesis~\cite{Luty:1992un,Covi:1996wh,Plumacher:1996kc,Pilaftsis:1997jf}
%%%%%%%%%%%%%%%%%%%%%%%%%%%%%%%%%%%%%%%%%%%%%%%%%%%%%%%%%%%%%%%%%%%%%%%%%%%%%
\footnote{Here we have used some {\it calligraphic} characters to denote the
relevant physical quantities in the basis where $M^{}_{\rm R}$ is diagonalized
by the unitary transformation made in Eq.~(\ref{4}). This basis is associated
with ${\cal N}^{}_i$ decays and thermal leptogenesis, and it is conceptually
different from the basis taken for the seesaw mechanism as can be seen below.}.
%%%%%%%%%%%%%%%%%%%%%%%%%%%%%%%%%%%%%%%%%%%%%%%%%%%%%%%%%%%%%%%%%%%%%%%%%%%%%%
To be more specific, the flavor-dependent CP-violating asymmetries of
${\cal N}^{}_i$ decays are given by
\begin{eqnarray}
\varepsilon^{}_{i \alpha} \hspace{-0.2cm} & \equiv & \hspace{-0.2cm}
\frac{\Gamma\big({\cal N}^{}_i \to \ell^{}_\alpha + H\big)
- \Gamma\big({\cal N}^{}_i \to \overline{\ell^{}_\alpha} +
\overline{H}\big)}{\displaystyle \sum_\alpha \left[\Gamma\big({\cal N}^{}_i \to
\ell^{}_\alpha + H\big) + \Gamma\big({\cal N}^{}_i \to \overline{\ell^{}_\alpha}
+ \overline{H}\big)\right]}
\nonumber \\
\hspace{-0.2cm} & = & \hspace{-0.2cm}
\frac{1}{8\pi \big({\cal Y}^\dagger_\nu
{\cal Y}^{}_\nu\big)^{}_{ii}} \sum_{j \neq i} \left\{ {\rm Im}
\left[\big({\cal Y}^*_\nu\big)^{}_{\alpha i}
\big({\cal Y}^{}_\nu\big)^{}_{\alpha j}
\big({\cal Y}^\dagger_\nu {\cal Y}^{}_\nu\big)^{}_{ij} \xi(x^{}_{ji})
+ \big({\cal Y}^*_\nu\big)^{}_{\alpha i}
\big({\cal Y}^{}_\nu\big)^{}_{\alpha j} \big({\cal Y}^\dagger_\nu
{\cal Y}^{}_\nu\big)^*_{ij} \zeta(x^{}_{ji}) \right]\right\} \; , \hspace{0.4cm}
\label{6}
%     (6)
\end{eqnarray}
where the Latin and Greek subscripts run respectively over $(1, 2, 3)$
and $(e, \mu, \tau)$, $x^{}_{ji} \equiv {\cal M}^2_j/{\cal M}^2_i$ are defined,
$\xi(x^{}_{ji}) = \sqrt{x^{}_{ji}} \left\{1 + 1/\left(1 - x^{}_{ji}\right)
+ \left(1 + x^{}_{ji}\right) \ln \left[x^{}_{ji} / \left(1 + x^{}_{ji}\right)
\right] \right\}$ and $\zeta(x^{}_{ji}) = 1/\left(1 - x^{}_{ji}\right)$ are
the loop functions. A net lepton-antilepton asymmetry can therefore result
from $\varepsilon^{}_{i \alpha}$ in the early Universe, and later on it
can be partly converted into a net baryon-antibaryon asymmetry
via the sphaleron interactions (see Ref.~\cite{DiBari:2021fhs} for a recent
review).

At this point it is worth remarking that the right-handed neutrino fields
$N^{}_{\alpha \rm R}$ have zero weak isospin and hypercharge, and hence
they have no coupling with the charged and neutral gauge bosons of the SM.
As a consequence, the mass eigenstates of heavy Majorana neutrinos obtained from
Eq.~(\ref{4}) do not participate in the weak charged-current interactions
of the SM,
\begin{eqnarray}
-{\cal L}^{}_{\rm cc} = \frac{g}{2} \hspace{0.05cm}
\overline{\ell^{}_{\rm L}} \hspace{0.05cm} \gamma^\mu \left(\sigma^{}_1
W^1_\mu + \sigma^{}_2 W^2_\mu\right) \ell^{}_{\rm L}
= \frac{g}{\sqrt 2} \hspace{0.05cm} \overline{l^{}_{\rm L}} \hspace{0.05cm}
\gamma^\mu \hspace{0.03cm} W^-_\mu \nu^{}_{\rm L} + {\rm h.c.} \; ,
\label{7}
%     (7)
\end{eqnarray}
where $g$ denotes the weak gauge coupling constant, $\sigma^{}_{1,2}$ represent
the first and second Pauli matrices, $W^\mu_{1,2}$ are two of the original
$\rm SU(2)^{}_{\rm L}$ gauge fields, and $W^\pm_\mu \equiv
\left(W^1_\mu \mp {\rm i} \hspace{0.03cm} W^2_\mu\right)/\sqrt{2}$ stand for
the fields of the physical charged gauge bosons $W^\pm$. But in the seesaw
framework we shall see that the expression of ${\cal L}^{}_{\rm cc}$ in
Eq.~(\ref{7}) will get modified, and the corresponding mass eigenstates of
three heavy Majorana neutrinos can definitely take part in the weak
charged-current interactions.

\subsection{The seesaw-associated basis}

We proceed to show that the canonical seesaw mechanism can ``formally" work before
spontaneous electroweak symmetry breaking but the corresponding mass eigenstates
of three heavy Majorana neutrinos are not exactly the same as ${\cal N}^{}_i$
(for $i = 1, 2, 3$) obtained above for the neutrino decays and thermal leptogenesis.
To clarify this important point, let us diagonalize the symmetric $6\times 6$
matrix in Eq.~(\ref{3}) in the following Autonne-Takagi way:
\begin{eqnarray}
\mathbb{U}^\dagger \left ( \begin{matrix} {\bf 0} & Y^{}_\nu \phi^{0*}
\cr Y^T_\nu \phi^{0*} & M^{}_{\rm R} \cr \end{matrix} \right )
\mathbb{U}^* = \left( \begin{matrix} D^{}_\nu & {\bf 0} \cr {\bf 0} &
D^{}_N \cr \end{matrix} \right) \; ,
\label{8}
%     (8)
\end{eqnarray}
where $\mathbb{U}$ is a $6\times 6$ unitary matrix, and the
diagonal and real matrices $D^{}_\nu$ and $D^{}_N$ are defined as
$D^{}_\nu \equiv {\rm Diag}\big\{m^{}_1, m^{}_2, m^{}_3\big\}$ and
$D^{}_N \equiv {\rm Diag}\big\{M^{}_1, M^{}_2, M^{}_3 \big\}$. Meanwhile,
the column vectors of left- and right-handed neutrino fields
$\big[\begin{matrix} \nu^{}_{\rm L} & (N^{}_{\rm R})^c\end{matrix}
\big]^T$ and $\big[\begin{matrix} (\nu^{}_{\rm L})^c & N^{}_{\rm R}
\end{matrix}\big]^T$ undergo the transformations
\begin{eqnarray}
\left[\begin{matrix} \nu^{}_{\rm L} \cr (N^{}_{\rm R})^c\end{matrix}
\right] \longrightarrow \hspace{0.1cm} \mathbb{U}^\dagger
\left[\begin{matrix} \nu^{}_{\rm L} \cr (N^{}_{\rm R})^c\end{matrix}
\right] \; , \quad
\left[\begin{matrix} (\nu^{}_{\rm L})^c \cr N^{}_{\rm R} \end{matrix}\right]
\longrightarrow \hspace{0.1cm} \mathbb{U}^T
\left[\begin{matrix} (\nu^{}_{\rm L})^c \cr N^{}_{\rm R} \end{matrix}\right] \; ,
\label{9}
%     (9)
\end{eqnarray}
such that the Lagrangian ${\cal L}^{}_\Lambda$ in Eq.~(\ref{3}) keeps unchanged
and thus its {\it overall} gauge symmetry is unbroken. Now that $Y^{}_\nu$
is dimensionless and $\phi^0$ has the same mass dimension as $M^{}_{\rm R}$, one
may argue that $m^{}_i$ should be the ``working" or ``virtual" mass parameters of
three light Majorana neutrinos as the electroweak gauge symmetry is unbroken at the
seesaw scale $\Lambda$. In comparison, $M^{}_i$ are essentially the true masses of
three heavy Majorana neutrinos in the existence of the $\phi^{0 (*)}$-mediated
neutrino Yukawa interactions. Along this line of thought, we find that it is
useful to decompose $\mathbb{U}$ into the product of three matrices,
\begin{eqnarray}
\mathbb{U} = \left( \begin{matrix}
I & {\bf 0} \cr {\bf 0} & U^{\prime}_0 \cr \end{matrix} \right)
\left( \begin{matrix} A & R \cr S & B \cr \end{matrix} \right)
\left( \begin{matrix} U^{}_0 & {\bf 0} \cr {\bf 0} & I \cr
\end{matrix} \right) \; ,
\label{10}
%     (10)
\end{eqnarray}
where the $3\times 3$ unitary matrix $U^\prime_0$ has been defined in Eq.~(\ref{4})
to primarily describe flavor mixing in the sterile (heavy) neutrino sector,
$U^{}_0$ denotes the other $3\times 3$ unitary matrix that is mainly responsible
for flavor mixing in the active (light) neutrino sector, while the $3\times 3$
matrices $A$, $B$, $R$ and $S$ signify the interplay between these two
sectors~\cite{Xing:2007zj,Xing:2011ur,Xing:2020ijf}. The unitarity of $\mathbb{U}$
assures
\begin{eqnarray}
A A^\dagger + R R^\dagger \hspace{-0.2cm} & = & \hspace{-0.2cm}
B B^\dagger + S S^\dagger = I \; ,
\nonumber \\
A S^\dagger + R B^\dagger \hspace{-0.2cm} & = & \hspace{-0.2cm}
A^\dagger R + S^\dagger B = {\bf 0} \; ,
\nonumber \\
A^\dagger A + S^\dagger S \hspace{-0.2cm} & = & \hspace{-0.2cm}
B^\dagger B + R^\dagger R = I \; .
\hspace{1.cm}
\label{11}
%     (11)
\end{eqnarray}
On the other hand, the arbitrary charged-lepton Yukawa coupling matrix
$Y^{}_l$ in Eq.~(\ref{3}) can be diagonalized by a bi-unitary transformation:
\begin{eqnarray}
U^{\dagger}_l \left(Y^{}_l \hspace{0.03cm} \phi^0\right) V^{}_l = D^{}_l \; ,
\quad l^\prime_{\rm L} = U^{\dagger}_l \hspace{0.02cm} l^{}_{\rm L} \; ,
\quad l^\prime_{\rm R} = V^{}_l \hspace{0.03cm} l^{}_{\rm R} \; ,
\label{12}
%     (12)
\end{eqnarray}
where $U^{}_l$ and $V^{}_l$ are unitary, $D^{}_l \equiv {\rm Diag}\big\{
m^{}_e , m^{}_\mu , m^{}_\tau\big\}$ stands for the ``working" or ``virtual"
masses of three charged leptons {\it before} spontaneous electroweak symmetry
breaking
%%%%%%%%%%%%%%%%%%%%%%%%%%%%%%%%%%%%%%%%%%%%%%%%%%%%%%%%%%%%%%%%%%%%%%%%%%%%
\footnote{Note that the scalar field $\phi^0$ in Eq.~(\ref{12}) carries a
hypercharge, and hence $D^{}_l$ cannot be simply understood as a diagonal
``mass" matrix. The physical meaning of $D^{}_l$ is actually vague in our
calculations which are mathematically exact and clear, so is the physical
meaning of $D^{}_\nu$ in Eq.~(\ref{8}). But this vagueness will automatically
disappear after spontaneous electroweak symmetry breaking, as can be
subsequently seen.},
%%%%%%%%%%%%%%%%%%%%%%%%%%%%%%%%%%%%%%%%%%%%%%%%%%%%%%%%%%%%%%%%%%%%%%%%%%%%
and $l^\prime = \big(\begin{matrix}
e & \mu & \tau \end{matrix}\big)^T$ is defined as the column vector of the
mass eigenstates of three charged leptons versus the column vector of their
flavor eigenstates $l = \big(\begin{matrix} l^{}_e & l^{}_\mu & l^{}_\tau
\end{matrix}\big)^T$. Substituting Eqs.~(\ref{8})---(\ref{10}) and (\ref{12})
into Eq.~(\ref{3}), we immediately arrive at
\begin{eqnarray}
-{\cal L}^{}_{\Lambda} = \overline{l^{\prime}_{\rm L}} \hspace{0.05cm} D^{}_l
\hspace{0.03cm} l^{\prime}_{\rm R} +
\frac{1}{2} \hspace{0.05cm} \overline{\nu^{\prime}_{\rm L}}
\hspace{0.05cm} D^{}_\nu (\nu^\prime_{\rm L})^c +
\frac{1}{2} \hspace{0.05cm} \overline{(N^{\prime}_{\rm R})^c} \hspace{0.05cm}
D^{}_N N^{\prime}_{\rm R} +
\overline{\nu^{}_{\rm L}} \hspace{0.05cm} Y^{}_l \hspace{0.03cm} l^{}_{\rm R}
\phi^+ - \overline{l^{}_{\rm L}} \hspace{0.05cm}
Y^{}_\nu N^{}_{\rm R} \phi^- + {\rm h.c.} \; ,
\label{13}
%     (13)
\end{eqnarray}
where $\nu^\prime = \big(\begin{matrix} \nu^{}_{1} & \nu^{}_{2} & \nu^{}_{3}
\end{matrix}\big)^T$ denotes the column vector of the {\it working} mass
eigenstates of three light Majorana neutrinos far above the electroweak
scale, and $N^\prime = \big(\begin{matrix} N^{}_{1} & N^{}_{2}
& N^{}_{3}\end{matrix}\big)^T$ stands for the column vectors of the mass
eigenstates of three heavy Majorana neutrinos relevant to the seesaw mechanism
at $\Lambda \gg v$. In this case the flavor eigenstates $\nu^{}_{\rm L}$ and
$N^{}_{\rm R}$ can be expressed in terms of the mass eigenstates
$\nu^\prime_{\rm L}$ and $N^\prime_{\rm R}$ or their charge-conjugated states
as follows:
\begin{eqnarray}
\nu^{}_{\rm L} = U \nu^\prime_{\rm L} + R \hspace{0.03cm} (N^\prime_{\rm R})^c \; ,
\quad
N^{}_{\rm R} = S^{\prime *} (\nu^\prime_{\rm L})^c
+ U^{\prime *} N^\prime_{\rm R} \; ,
\label{14}
%     (14)
\end{eqnarray}
where $U \equiv A U^{}_0$, $U^\prime \equiv U^\prime_0 B$ and
$S^\prime \equiv U^\prime_0 S U^{}_0$ are defined. Taking account of the Majorana
property of $\nu^{}_i$ and $N^{}_i$ (i.e., $\nu^c_i = \nu^{}_i$ and
$N^c_i = N^{}_i$~\cite{Majorana:1937vz} for $i = 1, 2, 3$), one simply obtains
$(N^\prime_{\rm R})^c = (N^{\prime c})^{}_{\rm L} = N^\prime_{\rm L}$
and $(\nu^\prime_{\rm L})^c = (\nu^{\prime c})^{}_{\rm R} = \nu^\prime_{\rm R}$.
One may then substitute the expression of $l^{}_{\rm L}$ in Eq.~(\ref{12}) and
that of $\nu^{}_{\rm L}$ in Eq. (\ref{14}) into the standard form of
${\cal L}^{}_{\rm cc}$ in Eq.~(\ref{7}) and get at
\begin{eqnarray}
-{\cal L}^{}_{\rm cc} = \frac{g}{\sqrt{2}} \hspace{0.1cm}
\overline{\big(\begin{matrix} e & \mu & \tau\end{matrix}\big)^{}_{\rm L}}
\hspace{0.1cm} \gamma^\mu \left[ U^{}_{\rm PMNS} \left( \begin{matrix} \nu^{}_{1}
\cr \nu^{}_{2} \cr \nu^{}_{3} \cr\end{matrix} \right)^{}_{\hspace{-0.08cm} \rm L}
+ R^{}_{\rm PMNS} \left(\begin{matrix} N^{}_{1} \cr N^{}_{2} \cr N^{}_{3}
\cr\end{matrix}\right)^{}_{\hspace{-0.08cm} \rm L} \hspace{0.05cm} \right]
W^-_\mu + {\rm h.c.} \; ,
\label{15}
%     (15)
\end{eqnarray}
where $U^{}_{\rm PMNS} = U^\dagger_l U$ is just the
Pontecorvo-Maki-Nakagawa-Sakata (PMNS) lepton flavor mixing matrix~\cite{Pontecorvo:1957cp,Maki:1962mu,Pontecorvo:1967fh} used to
describe the flavor oscillations of three active neutrinos, and
$R^{}_{\rm PMNS} = U^\dagger_l R$ is an analogue of $U^{}_{\rm PMNS}$ in the
seesaw mechanism which characterizes the strengths of weak charged-current
interactions for three heavy Majorana neutrinos.

Without loss of generality, one may choose a convenient flavor basis in which
the mass eigenstates of three charged leptons are identified with their
corresponding flavor eigenstates (i.e., $l^{}_{\rm L} = l^\prime_{\rm L}$,
or equivalently $U^{}_l = I$). In this case we are simply left with
$U^{}_{\rm PMNS} = U$ and $R^{}_{\rm PMNS} = R$, namely the effects of
lepton flavor mixing originate purely from the active and sterile Majorana
neutrino sectors and from the interplay between these two sectors. We
shall take advantage of this flavor basis in the following discussions
unless otherwise specified.

\subsection{Mismatch between the two bases}

Before discussing a mismatch between the mass eigenstates of heavy
Majorana neutrinos associated with thermal leptogenesis and those associated
with the seesaw mechanism, let us take a look at the flavor structures of
active and sterile neutrinos in the case that the electroweak gauge symmetry
is unbroken at $\Lambda$. First of all, a combination of Eqs.~(\ref{8}) and
(\ref{10}) allows us to immediately derive the exact seesaw relation between
the working masses of three light Majorana neutrinos and the real masses
of three heavy Majorana neutrinos:
\begin{eqnarray}
U D^{}_\nu U^T + R D^{}_N R^T = {\bf 0} \; ,
\label{16}
%     (16)
\end{eqnarray}
in which $U$ and $R$ are also correlated with each other via the unitarity
condition $U U^\dagger + R R^\dagger = I$. Note that $U = A U^{}_0$ holds,
where the unitary matrix $U^{}_0$ is primarily responsible for flavor mixing
of the three active neutrinos. So we find it useful to rewrite
Eq.~(\ref{16}) as
\begin{eqnarray}
U^{}_0 D^{}_\nu U^T_0 =  \left({\rm i} \hspace{0.03cm} A^{-1} R\right) D^{}_N
\left({\rm i} \hspace{0.03cm} A^{-1} R\right)^T \; ,
\label{17}
%     (17)
\end{eqnarray}
whose left- and right-hand sides are composed of the {\it derivational} and
{\it original} seesaw parameters, respectively. This point will become more
obvious when a complete Euler-like parametrization of the $6\times 6$ unitary
matrix $\mathbb{U}$ in Eq.~(\ref{10}) is adopted, as can be seen in
section~\ref{section 3}. Needless to say, the active-sterile flavor mixing
matrix $R$ essentially plays the role of the neutrino Yukawa coupling matrix
$Y^{}_\nu$ in the canonical seesaw framework,
\begin{eqnarray}
Y^{}_\nu \phi^{0 *} = R D^{}_N \left[ I - \left(B^{-1} S A^{-1} R\right)^T
\right] U^{\prime T} \; ;
\label{18}
%     (18)
\end{eqnarray}
and the right-handed Majorana neutrino mass matrix $M^{}_{\rm R}$ can be
reconstructed into the form
\begin{eqnarray}
M^{}_{\rm R} = U^\prime \left[ D^{}_N - \left(B^{-1} S A^{-1} R\right)
D^{}_N \left(B^{-1} S A^{-1} R\right)^T\right] U^{\prime T} \; .
\label{19}
%     (19)
\end{eqnarray}
Note that all the quantities in Eqs.~(\ref{18}) and (\ref{19}) belong to the
{\it original} seesaw parameters in the sense that they have nothing to do
with $D^{}_\nu$ and $U^{}_0$ --- the working masses and the primary flavor
mixing matrix of three light Majorana neutrinos which are {\it derived} from
the seesaw mechanism.

Now we turn to an unavoidable mismatch between the mass eigenstates of three
heavy Majorana neutrinos associated with the seesaw and leptogenesis mechanisms.
Eq.~(\ref{14}) tells us that the mass eigenstates $N^{\prime}_{\rm R}$ in
the seesaw basis can be expressed as
\begin{eqnarray}
N^\prime_{\rm R} = (U^{\prime *})^{-1} \left[ N^{}_{\rm R} - S^{\prime *}
(\nu^\prime_{\rm L})^c\right] =
(B^*)^{-1} \left[ {\cal N}^\prime_{\rm R} - U^{\prime T}_0 S^{\prime *}
(\nu^\prime_{\rm L})^c\right] \; ,
\label{20}
%     (20)
\end{eqnarray}
where Eq.~(\ref{4}) has been used to link $N^\prime_{\rm R}$ to
${\cal N}^\prime_{\rm R}$. To be more explicit, Eq.~(\ref{20}) means
\begin{eqnarray}
\left(\begin{matrix} N^{}_1 \cr N^{}_2 \cr N^{}_3 \end{matrix}\right) =
(B^*)^{-1} \left[ \left(\begin{matrix} {\cal N}^{}_1 \cr {\cal N}^{}_2
\cr {\cal N}^{}_3 \end{matrix}\right) - U^{\prime T}_0 S^{\prime *}
\left( \begin{matrix} \nu^{}_{1} \cr \nu^{}_{2} \cr \nu^{}_{3} \cr\end{matrix}
\right)\right] \; ,
\label{21}
%     (21)
\end{eqnarray}
from which the differences between $N^{}_i$ in the seesaw basis and
${\cal N}^{}_i$ (for $i = 1, 2, 3$) in the thermal leptogenesis basis can be
clearly seen. Similarly, a combination of Eqs.~(\ref{4}) and (\ref{19})
leads us to
\begin{eqnarray}
{\cal D}^{}_{\cal N} = B \left[ D^{}_N - \left(B^{-1} S A^{-1} R\right)
D^{}_N \left(B^{-1} S A^{-1} R\right)^T\right] B^{T} \; ,
\label{22}
%     (22)
\end{eqnarray}
from which one may easily see the difference between ${\cal D}^{}_N$ and
$D^{}_N$. Although $N^\prime_{\rm R}$ (or $D^{}_N$) and ${\cal N}^\prime_{\rm R}$
(or ${\cal D}^{}_{\cal N}$) would exactly coincide with each other if the Yukawa
coupling matrix $Y^{}_\nu$ (or equivalently, $R$ or $S$) were switched off, such
a coincidence would make no sense because both the seesaw and leptogenesis
mechanisms would fail in this special case. In the presence of the neutrino Yukawa
interactions, thermal leptogenesis may take effect via the CP-violating and
out-of-equilibrium decays of heavy Majorana neutrinos into the leptonic and
Higgs doublets, while the seesaw mechanism can ``formally" work with the help
of an interplay between the active and sterile neutrino fields coupled only to
the neutral component of the Higgs doublet. That is the key reason why there
is an inevitable mismatch between the seesaw- and leptogenesis-associated bases
for heavy Majorana neutrinos before spontaneous electroweak symmetry breaking.

\subsection{After gauge symmetry breaking}

So far we have made some proper transformations of the charged lepton
and neutrino fields in the flavor space to obtain their respective working or
true mass eigenstates. All such unitary flavor basis transformations are
completely reversible, and hence they do not affect the gauge invariance of
${\cal L}^{}_\Lambda$ at the seesaw scale. As already shown in Eqs.~(\ref{20})
and (\ref{22}), a seeable mismatch between $N^\prime_{\rm R}$ and
${\cal N}^\prime_{\rm R}$ or between $D^{}_N$ and ${\cal D}^{}_N$ results
from the fact that the working seesaw mechanism itself is
only associated with the neutral component of the Higgs doublet while the
heavy Majorana neutrino decays and thermal leptogenesis at the seesaw scale
$\Lambda$ are associated with the whole Higgs doublet. This unavoidable
mismatch deserves to be conceptually clarified as we have done, because it
is an intrinsic issue of the seesaw and leptogenesis mechanisms.

It is now straightforward to prove that the {\it formal} seesaw mechanism
far above the electroweak scale will become {\it real} after the Higgs
potential of the SM is minimized at
$\langle H\rangle \equiv \langle 0|H|0\rangle = v/\sqrt{2}$ with a
special direction characterized by $\langle \phi^\pm\rangle = 0$ and
$\langle \phi^0\rangle = v/\sqrt{2}$, by which the electroweak gauge
symmetry is spontaneously broken and thus all the particles coupled to the
Higgs field acquire their nonzero masses. In this case the Lagrangian
in Eq.~(\ref{3}) can be simplified to a more popular form,
\begin{eqnarray}
-{\cal L}^{\prime}_{\Lambda} = \overline{l^{}_{\rm L}} \hspace{0.05cm}
M^{}_l \hspace{0.03cm} l^{}_{\rm R} + \frac{1}{2} \hspace{0.05cm}
\overline{\big[\begin{matrix} \nu^{}_{\rm L} & (N^{}_{\rm R})^c\end{matrix}
\big]} \left(\begin{matrix} {\bf 0} & M^{}_{\rm D} \cr
M^T_{\rm D} & M^{}_{\rm R} \end{matrix}\right)
\left[\begin{matrix} (\nu^{}_{\rm L})^c \cr N^{}_{\rm R} \end{matrix}\right]
+ {\rm h.c.} \; ,
\label{23}
%     (23)
\end{eqnarray}
where $M^{}_l \equiv Y^{}_l \langle \phi^0\rangle = Y^{}_l v/\sqrt{2}$ denotes
the charged lepton mass matrix, and $M^{}_{\rm D} \equiv Y^{}_\nu \langle
\phi^0\rangle = Y^{}_\nu v/\sqrt{2}$ is usually referred to as the Dirac
neutrino mass matrix. The expression of $M^{}_{\rm D}$ in terms of the seesaw
parameters can be directly read off from Eq.~(\ref{18}), namely
\begin{eqnarray}
M^{}_{\rm D} = R D^{}_N \left[ I - \left(B^{-1} S A^{-1} R\right)^T
\right] U^{\prime T} \; .
\label{24}
%     (24)
\end{eqnarray}
We find that the exact seesaw formula obtained in Eq.~(\ref{16}) and the
analytical results obtained in Eqs.~(\ref{19})---(\ref{22}) formally keep
unchanged after spontaneous gauge symmetry breaking, but they are now
subject to the electroweak scale. In other words, the electroweak symmetry
breaking itself does not really affect the flavor structures of the seesaw
mechanism. This observation implies that it is possible to determine or
constrain some of the original seesaw-associated flavor parameters in some
low-energy neutrino experiments, after the radiative corrections to such
parameters are properly taken into account with the help of the relevant
renormaliztion-group equations (RGEs) between a superhigh seesaw scale and the
electroweak scale~\cite{Ohlsson:2013xva}.

Note that the exact seesaw formula obtained in Eq.~(\ref{16})
can be simplified to the more popular form in the leading-order approximations
of Eqs.~(\ref{19}) and (\ref{24}).
That is, $M^{}_{\rm R} \simeq U^\prime D^{}_N U^{\prime T}$
and $M^{}_{\rm D} \simeq R D^{}_N U^{\prime T}$, so the {\it effective}
mass matrix for three active Majorana neutrinos is given by
\begin{eqnarray}
M^{}_\nu \equiv U^{}_0 D^{}_\nu U^T_0 \simeq -R D^{}_N R^T \simeq
-M^{}_{\rm D} M^{-1}_{\rm R} M^{T}_{\rm D} \; ,
\label{25}
%     (25)
\end{eqnarray}
where $A \simeq B \simeq I$ has been assumed (i.e., $U \simeq U^{}_0$ holds
in the neglect of the non-unitary effects characterized by $A \neq I$). In
this approximation the effective Majorana mass term for three
active neutrinos at low energies turns out to be
\begin{eqnarray}
-{\cal L}^{}_\nu = \frac{1}{2} \hspace{0.05cm} \overline{\nu^{}_{\rm L}}
\hspace{0.07cm} M^{}_\nu (\nu^{}_{\rm L})^c + {\rm h.c.} =
\frac{1}{2} \hspace{0.05cm} \overline{\nu^{\prime}_{\rm L}}
\hspace{0.05cm} D^{}_\nu (\nu^\prime_{\rm L})^c + {\rm h.c.} \; ,
\label{26}
%     (26)
\end{eqnarray}
where the column vector of the light neutrino mass eigenstates
$\nu^\prime_{\rm L}$ has already been defined below Eq.~(\ref{13}), and
the physical meaning of $D^{}_\nu$ as the diagonal Majorana neutrino mass
matrix becomes definite and obvious.

\section{How small is the mismatch?}
\label{section 3}

\subsection{An Euler-like parametrization}

To clearly see how small the difference between $N^\prime_{\rm R}$ (or $D^{}_N$)
and ${\cal N}^\prime_{\rm R}$ (or ${\cal D}^{}_{\cal N}$) is expected to be, let
us follow Refs.~\cite{Xing:2007zj,Xing:2011ur,Xing:2020ijf} to make an Euler-like
parametrization of the $6\times 6$ unitary matrix $\mathbb{U}$ in Eq.~(\ref{10}).
First of all we introduce fifteen $6\times 6$ Euler-like unitary
matrices of the form $O^{}_{ij}$ (for $1 \leq i < j \leq 6$): its $(i, i)$ and
$(j, j)$ entries are both identical to $c^{}_{ij} \equiv \cos\theta^{}_{ij}$
with $\theta^{}_{ij}$ being a flavor mixing angle and lying in the first quadrant,
its other four diagonal elements are all equal to one, its $(i, j)$ and $(j, i)$
entries are given respectively by $\hat{s}^{*}_{ij} \equiv e^{-{\rm i}\delta^{}_{ij}}
\sin\theta^{}_{ij}$ and $-\hat{s}^{}_{ij} \equiv -e^{{\rm i}\delta^{}_{ij}}
\sin\theta^{}_{ij}$ with $\delta^{}_{ij}$ being a CP-violating phase, and its
other off-diagonal elements are all equal to zero. These matrices are then grouped
in the following way to respectively describe the {\it active} flavor sector,
the {\it sterile} flavor sector and the {\it interplay} between these two sectors:
\begin{eqnarray}
\left( \begin{matrix} U^{}_0 & 0 \cr 0 & I \cr \end{matrix} \right)
\hspace{-0.2cm} & = & \hspace{-0.2cm} O^{}_{23} O^{}_{13} O^{}_{12} \; ,
\nonumber \\
\left( \begin{matrix} I & 0 \cr 0 & U^{\prime}_0 \cr \end{matrix} \right)
\hspace{-0.2cm} & = & \hspace{-0.2cm} O^{}_{56} O^{}_{46} O^{}_{45} \; ,
\nonumber \\
\left( \begin{matrix} A & R \cr S & B \cr \end{matrix} \right)
\hspace{-0.2cm} & = & \hspace{-0.2cm} O^{}_{36} O^{}_{26} O^{}_{16}
O^{}_{35} O^{}_{25} O^{}_{15} O^{}_{34} O^{}_{24} O^{}_{14} \; , \hspace{0.7cm}
\label{27}
%     (27)
\end{eqnarray}
where the pattern of $U^{}_0$ is quite similar to the standard parametrization
of a unitary PMNS matrix as advocated by the Particle Data
Group~\cite{ParticleDataGroup:2022pth}
%%%%%%%%%%%%%%%%%%%%%%%%%%%%%%%%%%%%%%%%%%%%%%%%%%%%%%%%%%%%%%%%%%%%%%%%%%%%%%%
\footnote{When $U^{}_0$ is applied to the phenomenology of neutrino physics
in the basis of $U^{}_l = I$, it is the phase parameter
$\delta \equiv \delta^{}_{13} - \delta^{}_{12} - \delta^{}_{23}$ that
characterizes the strength of CP violation in neutrino oscillations.},
%%%%%%%%%%%%%%%%%%%%%%%%%%%%%%%%%%%%%%%%%%%%%%%%%%%%%%%%%%%%%%%%%%%%%%%%%%%%%%%
\begin{eqnarray}
U^{}_0 \hspace{-0.2cm} & = & \hspace{-0.2cm}
\left( \begin{matrix} c^{}_{12} c^{}_{13} & \hat{s}^*_{12}
c^{}_{13} & \hat{s}^*_{13} \cr
-\hat{s}^{}_{12} c^{}_{23} -
c^{}_{12} \hat{s}^{}_{13} \hat{s}^*_{23} & c^{}_{12} c^{}_{23} -
\hat{s}^*_{12} \hat{s}^{}_{13} \hat{s}^*_{23} & c^{}_{13}
\hat{s}^*_{23} \cr
\hat{s}^{}_{12} \hat{s}^{}_{23} - c^{}_{12}
\hat{s}^{}_{13} c^{}_{23} & -c^{}_{12} \hat{s}^{}_{23} -
\hat{s}^*_{12} \hat{s}^{}_{13} c^{}_{23} & c^{}_{13} c^{}_{23}
\cr \end{matrix} \right) \; , \hspace{0.4cm}
\label{28}
%     (28)
\end{eqnarray}
and the expression of $U^\prime_0$ can be directly read off from that of $U^{}_0$
with the subscript replacements $12 \leftrightarrow 45$, $13 \leftrightarrow 46$
and $23 \leftrightarrow 56$ for the three rotation angles and three
CP-violating phases. The explicit expressions of $A$, $B$, $R$ and $S$ in
terms of $c^{}_{ij}$ and $\hat{s}^{}_{ij}$ (for $i = 1, 2, 3$ and
$j = 4, 5, 6$) are rather lengthy, and hence they are listed in
Eqs.~(\ref{A.1}) and (\ref{A.2}) in Appendix~\ref{A} for the same of simplicity.
Among the four active-sterile flavor mixing matrices, only $A$ and $R$ affect
the physical processes in which the light and heavy Majorana neutrinos take
part, as can be seen from Eq.~(\ref{15}). As both $U = A U^{}_0$ and $R$
appear in ${\cal L}^{}_{\rm cc}$ in the chosen flavor basis (i.e., $U^{}_l = I$),
three of the nice CP-violating phases (or their combinations) of $A$
and $R$ can always be rotated away by properly redefining the phases of
three charged lepton fields~\cite{Endoh:2000hc,Branco:2001pq}.

The PMNS matrix $U$ is obviously non-unitary because of $UU^\dagger =
A A^\dagger = I - RR^\dagger \neq I$, but its deviation from exact unitarity
(i.e., from $U^{}_0$) is found to be very small. A detailed and careful
analysis of currently available electroweak precision measurements and
neutrino oscillation data has put a stringent constraint on the non-unitarity
of $U$ --- the latter is below or far below
${\cal O}(10^{-2})$~\cite{Antusch:2006vwa,
Antusch:2009gn,Blennow:2016jkn,Hu:2020oba,Wang:2021rsi}.
This result implies that the deviation of $A A^\dagger$ from $I$ ought to
be smaller than ${\cal O}(10^{-2})$, and thus the nine active-sterile
flavor mixing angles in $R$ should be smaller than ${\cal O}(10^{-1})$. The
advantage of such a phenomenological observation is that $U \simeq U^{}_0$
can be a quite reliable approximation in most cases, but its disadvantage
is that an experimental exploration of the seesaw-induced non-unitary effects
of $U$ at low energies will be rather challenging.

\subsection{Smallness of the mismatch}

Eq.~(\ref{20}) tells us that a difference between the mass eigenstates of
three heavy Majorana neutrinos associated with the seesaw mechanism
(i.e., $N^\prime_{\rm R}$) and those associated with thermal leptogenesis
(i.e., ${\cal N}^\prime_{\rm R}$) is mainly characterized by the deviation
of $(B^*)^{-1}$ from the identity matrix $I$. With the help of Eq.~(\ref{A.2}),
we arrive at
\begin{eqnarray}
(B^*)^{-1} \hspace{-0.2cm} & = & \hspace{-0.2cm}
\left( \begin{matrix} c^{-1}_{14} c^{-1}_{24} c^{-1}_{34} & 0 & 0
\cr \vspace{-0.45cm} \cr
\begin{array}{l}
+ \hat{t}^{}_{14} c^{-1}_{24} c^{-1}_{34} \hat{t}^{*}_{15} +
\hat{t}^{}_{24} c^{-1}_{34} c^{-1}_{24} \hat{t}^{*}_{25} \\
+ \hat{t}^{}_{34} c^{-1}_{15} c^{-1}_{25} \hat{t}^{*}_{35} \end{array} &
c^{-1}_{15} c^{-1}_{25} c^{-1}_{35} & 0
\cr \vspace{-0.45cm} \cr
\begin{array}{l}
+ \hat{t}^{}_{14} c^{-1}_{24} c^{-1}_{34} c^{-1}_{24} \hat{t}^{*}_{16}
+ \hat{t}^{}_{24} c^{-1}_{34} \hat{t}^{}_{15} \hat{t}^{*}_{25} \hat{t}^{*}_{16} \\
+ \hat{t}^{}_{24} c^{-1}_{34} c^{-1}_{25} c^{-1}_{16} \hat{t}^{*}_{26}
+ \hat{t}^{}_{34} \hat{t}^{}_{15} c^{-1}_{25} \hat{t}^{*}_{35} \hat{t}^{*}_{16} \\
+ \hat{t}^{}_{34} c^{-1}_{35} c^{-1}_{16} c^{-1}_{26} \hat{t}^{*}_{36}
+ \hat{t}^{}_{34} \hat{t}^{}_{25} \hat{t}^{*}_{35} c^{-1}_{16} \hat{t}^{*}_{26}
\end{array} &
\begin{array}{l}
+ \hat{t}^{}_{15} c^{-1}_{25} c^{-1}_{35} \hat{t}^{*}_{16}
+ \hat{t}^{}_{25} c^{-1}_{35} c^{-1}_{16} \hat{t}^{*}_{26} \\
+ \hat{t}^{}_{35} c^{-1}_{16} c^{-1}_{26} \hat{t}^{*}_{36} \end{array} &
c^{-1}_{16} c^{-1}_{26} c^{-1}_{36}
\cr \end{matrix} \right) \; \hspace{0.2cm}
\nonumber \\
\hspace{-0.2cm} & \simeq & \hspace{-0.2cm}
I + \left(\begin{matrix}
\displaystyle\frac{1}{2} \left(s^2_{14} + s^2_{24} + s^2_{34}\right) & 0 & 0 \cr
\hat{s}^{}_{14} \hat{s}^*_{15} + \hat{s}^{}_{24} \hat{s}^*_{25} +
\hat{s}^{}_{34} \hat{s}^*_{35} & \displaystyle\frac{1}{2}
\left(s^2_{15} + s^2_{25} + s^2_{35} \right) & 0 \cr
\hat{s}^{}_{14} \hat{s}^*_{16} + \hat{s}^{}_{24} \hat{s}^*_{26} +
\hat{s}^{}_{34} \hat{s}^*_{36} &
~\hat{s}^{}_{15} \hat{s}^*_{16} + \hat{s}^{}_{25} \hat{s}^*_{26} +
\hat{s}^{}_{35} \hat{s}^*_{36}~ & \displaystyle\frac{1}{2}
\left(s^2_{16} + s^2_{26} + s^2_{36}\right) \cr \end{matrix}\right) \; ,
\label{29}
%     (29)
\end{eqnarray}
where $\hat{t}^{}_{ij} \equiv e^{{\rm i} \delta^{}_{ij}} \tan\theta^{}_{ij}$
is defined, and all the terms of ${\cal O}(s^4_{ij})$ or smaller have been
omitted from the second equation as an excellent approximation due to the
smallness of $\theta^{}_{ij}$ (for $i = 1, 2, 3$ and $j = 4, 5, 6$). We see that
$(B^*)^{-1}$ is also a lower triangular matrix like $B$ itself. On the other
hand, the factor $U^{\prime T}_0 S^{\prime *}$ appearing in Eq.~(\ref{20})
can be explicitly expressed as follows:
\begin{eqnarray}
U^{\prime T}_0 S^{\prime *} = B^T S^* U^*_0 \simeq
-\left(\begin{matrix} \hat{s}^{*}_{14} & \hat{s}^{*}_{24} & \hat{s}^{*}_{34} \cr
\hat{s}^{*}_{15} & \hat{s}^{*}_{25} & \hat{s}^{*}_{35} \cr
\hat{s}^{*}_{16} & \hat{s}^{*}_{26} & \hat{s}^{*}_{36} \cr \end{matrix}\right)
U^*_0 \; ,
\label{30}
%     (30)
\end{eqnarray}
where Eq.~(\ref{A.2}) has been used, and the terms of ${\cal O}(s^3_{ij})$ or
smaller have been omitted from the second equation as a very good approximation.
Now we conclude that the heavy Majorana neutrino mass basis
$N^\prime_{\rm R}$ is identical to ${\cal N}^\prime_{\rm R}$ up to the
accuracy of ${\cal O}(s^2_{ij})$, but it contains a small contribution of
${\cal O}(s^{}_{ij})$ from the light Majorana neutrino mass basis
$(\nu^\prime_{\rm L})^c$ in the seesaw framework. Since the magnitudes of
$\theta^{}_{ij}$ (for $i = 1, 2, 3$ and $j = 4, 5, 6$) are highly suppressed
in a realistic seesaw model with little fine-tuning, the mismatch between
$N^\prime_{\rm R}$ and ${\cal N}^\prime_{\rm R}$ is expected to be negligible
in most cases.

Let us proceed to examine how small the difference between $D^{}_N$ and
${\cal D}^{}_{\cal N}$ in Eq.~(\ref{21}) can be. First of all, Eq.~(\ref{A.2})
allows us to make the approximation
\begin{eqnarray}
B \simeq I - \left( \begin{matrix} \displaystyle\frac{1}{2} \left(s^2_{14} +
s^2_{24} + s^2_{34}\right) & 0 & 0 \cr \vspace{-0.4cm} \cr
\hat{s}^{*}_{14} \hat{s}^{}_{15} + \hat{s}^{*}_{24} \hat{s}^{}_{25}
+ \hat{s}^{*}_{34} \hat{s}^{}_{35} & \displaystyle\frac{1}{2}
\left(s^2_{15} + s^2_{25} + s^2_{35}\right)
& 0 \cr \vspace{-0.4cm} \cr
\hat{s}^{*}_{14} \hat{s}^{}_{16} + \hat{s}^{*}_{24}
\hat{s}^{}_{26} + \hat{s}^{*}_{34} \hat{s}^{}_{36} & ~\hat{s}^{*}_{15}
\hat{s}^{}_{16} + \hat{s}^{*}_{25} \hat{s}^{}_{26} + \hat{s}^{*}_{35}
\hat{s}^{}_{36}~ & \displaystyle\frac{1}{2} \left(s^2_{16} + s^2_{26} +
s^2_{36}\right) \cr \end{matrix} \right) \; ,
\label{31}
%     (31)
\end{eqnarray}
where the terms of ${\cal O}(s^4_{ij})$ or smaller have been omitted.
Secondly, we obtain
\begin{eqnarray}
A^{-1} R \hspace{-0.2cm} & = & \hspace{-0.2cm}
\left( \begin{matrix} \hat{t}^{*}_{14} & c^{-1}_{14} \hat{t}^{*}_{15} &
c^{-1}_{14} c^{-1}_{15} \hat{t}^{*}_{16}
\cr \vspace{-0.45cm} \cr
c^{-1}_{14} \hat{t}^{*}_{24} &
\hat{t}^{}_{14} \hat{t}^{*}_{15} \hat{t}^{*}_{24} + c^{-1}_{15} c^{-1}_{24}
\hat{t}^{*}_{25} &
\begin{array}{l}
+ \hat{t}^{}_{14} c^{-1}_{15} \hat{t}^{*}_{16} \hat{t}^{*}_{24}
+ \hat{t}^{}_{15} \hat{t}^{*}_{16} c^{-1}_{24} \hat{t}^{*}_{25} \\
+ c^{-1}_{16} c^{-1}_{24} c^{-1}_{25} \hat{t}^{*}_{26} \end{array}
\cr \vspace{-0.45cm} \cr
c^{-1}_{14} c^{-1}_{24} \hat{t}^{*}_{34} &
\begin{array}{l}
+ \hat{t}^{}_{14} \hat{t}^{*}_{15} c^{-1}_{24} \hat{t}^{*}_{34}
+ c^{-1}_{15} \hat{t}^{}_{24} \hat{t}^{*}_{25} \hat{t}^{*}_{34} \\
+ c^{-1}_{15} c^{-1}_{25} c^{-1}_{34} \hat{t}^{*}_{35}
\end{array} &
\begin{array}{l}
+ \hat{t}^{}_{14} c^{-1}_{15} \hat{t}^{*}_{16} c^{-1}_{24} \hat{t}^{*}_{34}
+ \hat{t}^{}_{15} \hat{t}^{*}_{16} \hat{t}^{}_{24} \hat{t}^{*}_{25}
\hat{t}^{*}_{34} \\
+ \hat{t}^{}_{15} \hat{t}^{*}_{16} c^{-1}_{25} c^{-1}_{34} \hat{t}^{*}_{35}
+ c^{-1}_{16} \hat{t}^{}_{24} c^{-1}_{25} \hat{t}^{*}_{26} \hat{t}^{*}_{34} \\
+ c^{-1}_{16} \hat{t}^{}_{25} \hat{t}^{*}_{26} c^{-1}_{34} \hat{t}^{*}_{35}
+ c^{-1}_{16} c^{-1}_{26} c^{-1}_{34} c^{-1}_{35} \hat{t}^{*}_{36}
\end{array}
\cr \end{matrix} \right) \; \hspace{0.2cm}
\nonumber \\
\hspace{-0.2cm} & \simeq & \hspace{-0.2cm}
\left(\begin{matrix} \hat{s}^*_{14} & \hat{s}^*_{15} & \hat{s}^*_{16} \cr
\hat{s}^*_{24} & \hat{s}^*_{25} & \hat{s}^*_{26} \cr
\hat{s}^*_{34} & \hat{s}^*_{35} & \hat{s}^*_{36} \cr \end{matrix}\right)
\label{32}
%     (32)
\end{eqnarray}
from Eq.~(\ref{A.1}), where the terms of ${\cal O}(s^3_{ij})$ or smaller
have been neglected in the second equation as a reasonably good
approximation. The exact expression of $B^{-1} S$ can be directly read off
from that of $-\left(A^{-1} R\right)^*$ with the help of Eq.~(\ref{32})
by making the subscript replacements $15 \leftrightarrow 24$,
$16 \leftrightarrow 34$ and $26 \leftrightarrow 35$, so can its approximate
expression. As a consequence,
\begin{eqnarray}
B^{-1} S A^{-1} R \simeq
-\left( \begin{matrix} s^2_{14} + s^2_{24} + s^2_{34} &
~\hat{s}^{}_{14} \hat{s}^{*}_{15} + \hat{s}^{}_{24} \hat{s}^{*}_{25}
+ \hat{s}^{}_{34} \hat{s}^{*}_{35}~ &
\hat{s}^{}_{14} \hat{s}^{*}_{16} + \hat{s}^{}_{24} \hat{s}^{*}_{26}
+ \hat{s}^{}_{34} \hat{s}^{*}_{36} \cr
\hat{s}^{*}_{14} \hat{s}^{}_{15} + \hat{s}^{*}_{24} \hat{s}^{}_{25}
+ \hat{s}^{*}_{34} \hat{s}^{}_{35} & s^2_{15} + s^2_{25} + s^2_{35} &
\hat{s}^{}_{15} \hat{s}^{*}_{16} + \hat{s}^{}_{25}
\hat{s}^{*}_{26} + \hat{s}^{}_{35} \hat{s}^{*}_{36} \cr
\hat{s}^{*}_{14} \hat{s}^{}_{16} + \hat{s}^{*}_{24} \hat{s}^{}_{26}
+ \hat{s}^{*}_{34} \hat{s}^{}_{36} &
~\hat{s}^{*}_{15} \hat{s}^{}_{16} + \hat{s}^{*}_{25} \hat{s}^{}_{26}
+ \hat{s}^{*}_{35} \hat{s}^{}_{36}~ &
s^2_{16} + s^2_{26} + s^2_{36} \cr \end{matrix} \right) \;
\label{33}
%     (33)
\end{eqnarray}
holds in the same approximation as made above. This result implies that
${\cal D}^{}_{\cal N}$ and $D^{}_N$ are identical to each other up to
the accuracy of ${\cal O}(s^2_{ij})$, simply because
on the right-hand side of Eq.~(\ref{22}) the second term is
suppressed in magnitude to ${\cal O}(s^4_{ij})$ as compared with
the first term.

It is worth remarking that our above analytical approximations are more
or less subject to the canonical
seesaw mechanism at an energy scale far above the electroweak scale, and
thus the mismatch between $N^\prime_{\rm R}$ (or $D^{}_N$)
and ${\cal N}^\prime_{\rm R}$ (or ${\cal D}^{}_{\cal N}$) is very small.
This situation will change when the low-scale seesaw and leptogenesis
scenarios, in which a mismatch between the two sets of mass bases for
heavy Majorana neutrinos is crucial, are taken into account (see, e.g.,
Refs.~\cite{Canetti:2012kh,Drewes:2019byd}).

\subsection{Determination of $D^{}_\nu$ and $U^{}_0$}

As already shown in Eq.~(\ref{17}), the nine effective flavor parameters
of three light Majorana neutrinos in $D^{}_\nu$ and $U^{}_0$ (i.e., three
effective masses, three flavor mixing angles and three CP-violating phases)
can be expressed in terms of the eighteen seesaw parameters hidden in $A$,
$R$ and $D^{}_N$ (i.e., three heavy Majorana neutrino masses, nine
active-sterile flavor mixing angles and six CP-violating phases). It is
obvious that all the derivational seesaw parameters on the left-hand side
of Eq.~(\ref{17}) would vanish if $R \propto Y^{}_\nu$ were switched off.
So this equation provides an unambiguous way to determine the light degrees
of freedom from the heavy degrees of freedom in the seesaw framework.

To be more specific, the six independent elements of the effective Majorana
neutrino mass matrix $M^{}_\nu \equiv U^{}_0 D^{}_\nu U^T_0$ are given as
follows:
\begin{eqnarray}
\big(M^{}_\nu\big)^{}_{11} \hspace{-0.2cm} & = & \hspace{-0.2cm}
m^{}_1 c^2_{12} c^2_{13} + m^{}_2 \hat{s}^{* 2}_{12} c^2_{13} +
m^{}_3 \hat{s}^{* 2}_{13} \; ,
\nonumber \\
\big(M^{}_\nu\big)^{}_{12} \hspace{-0.2cm} & = & \hspace{-0.2cm}
-m^{}_1 c^{}_{12} c^{}_{13} \big(\hat{s}^{}_{12} c^{}_{23} + c^{}_{12}
\hat{s}^{}_{13} \hat{s}^*_{23}\big) + m^{}_2 \hat{s}^*_{12} c^{}_{13}
\big(c^{}_{12} c^{}_{23} - \hat{s}^*_{12} \hat{s}^{}_{13}
\hat{s}^*_{23}\big) + m^{}_3 c^{}_{13} \hat{s}^{*}_{13}
\hat{s}^{*}_{23} \; , \hspace{0.3cm}
\nonumber \\
\big(M^{}_\nu\big)^{}_{13} \hspace{-0.2cm} & = & \hspace{-0.2cm}
m^{}_1 c^{}_{12} c^{}_{13} \big(\hat{s}^{}_{12} \hat{s}^{}_{23} - c^{}_{12}
\hat{s}^{}_{13} c^{}_{23}\big) - m^{}_2 \hat{s}^*_{12} c^{}_{13}
\big(c^{}_{12} \hat{s}^{}_{23} + \hat{s}^*_{12} \hat{s}^{}_{13}
c^{}_{23}\big) + m^{}_3 c^{}_{13} \hat{s}^{*}_{13} c^{}_{23} \; ,
\nonumber \\
\big(M^{}_\nu\big)^{}_{22} \hspace{-0.2cm} & = & \hspace{-0.2cm}
m^{}_1 \big(\hat{s}^{}_{12} c^{}_{23} + c^{}_{12} \hat{s}^{}_{13}
\hat{s}^*_{23}\big)^2 + m^{}_2 \big(c^{}_{12} c^{}_{23} -
\hat{s}^*_{12} \hat{s}^{}_{13} \hat{s}^*_{23}\big)^2 +
m^{}_3 c^2_{13} \hat{s}^{* 2}_{23} \; ,
\nonumber \\
\big(M^{}_\nu\big)^{}_{23} \hspace{-0.2cm} & = & \hspace{-0.2cm}
-m^{}_1 \big(\hat{s}^{}_{12} c^{}_{23} + c^{}_{12} \hat{s}^{}_{13}
\hat{s}^*_{23}\big) \big(\hat{s}^{}_{12} \hat{s}^{}_{23} - c^{}_{12}
\hat{s}^{}_{13} c^{}_{23}\big)
\nonumber \\
\hspace{-0.2cm} & & \hspace{-0.2cm}
- m^{}_2 \big(c^{}_{12} c^{}_{23} -
\hat{s}^*_{12} \hat{s}^{}_{13} \hat{s}^*_{23}\big)
\big(c^{}_{12} \hat{s}^{}_{23} + \hat{s}^*_{12} \hat{s}^{}_{13}
c^{}_{23}\big) + m^{}_3 c^{2}_{13} c^{}_{23} \hat{s}^{*}_{23} \; ,
\nonumber \\
\big(M^{}_\nu\big)^{}_{33} \hspace{-0.2cm} & = & \hspace{-0.2cm}
m^{}_1 \big(\hat{s}^{}_{12} \hat{s}^{}_{23} - c^{}_{12} \hat{s}^{}_{13}
c^{}_{23}\big)^2 + m^{}_2 \big(c^{}_{12} \hat{s}^{}_{23} +
\hat{s}^*_{12} \hat{s}^{}_{13} c^{}_{23}\big)^2 +
m^{}_3 c^2_{13} c^{2}_{23} \; .
\label{34}
%     (34)
\end{eqnarray}
On the other hand, Eq.~({\ref{17}) tells us that these six matrix elements
can originally be determined by $M^{}_\nu = -\big(A^{-1} R\big) D^{}_N
\big(A^{-1} R\big)^T$ thanks to the exact seesaw relation bridging the
big gap between the light and heavy Majorana neutrinos. With the help of the
explicit expression of $A^{-1} R$ given in Eq.~(\ref{32}), it is straightforward
to obtain the expressions for the elements of $M^{}_\nu$ in terms of $M^{}_i$,
$\theta^{}_{ij}$ and $\delta^{}_{ij}$ (for $i = 1, 2, 3$ and $j = 4, 5, 6$).
Instead of presenting the exact analytical results, which are rather lengthy
and hence less instructive, here we make the leading-order approximations for
the expressions of $A$ and $R$ given in Eq.~(\ref{A.1}) and then arrive at
\begin{eqnarray}
\big(M^{}_\nu\big)^{}_{11} \hspace{-0.2cm} & \simeq & \hspace{-0.2cm}
-\left[M^{}_1 \hat{s}^{* 2}_{14} + M^{}_2 \hat{s}^{* 2}_{15} +
M^{}_3 \hat{s}^{* 2}_{16}\right]  \; ,
\nonumber \\
\big(M^{}_\nu\big)^{}_{12} \hspace{-0.2cm} & \simeq & \hspace{-0.2cm}
-\left[M^{}_1 \hat{s}^{*}_{14} \hat{s}^{*}_{24} +
M^{}_2 \hat{s}^{*}_{15} \hat{s}^{*}_{25} +
M^{}_3 \hat{s}^{*}_{16} \hat{s}^{*}_{26}\right] \; ,
\nonumber \\
\big(M^{}_\nu\big)^{}_{13} \hspace{-0.2cm} & \simeq & \hspace{-0.2cm}
-\left[M^{}_1 \hat{s}^{*}_{14} \hat{s}^{*}_{34} +
M^{}_2 \hat{s}^{*}_{15} \hat{s}^{*}_{35} +
M^{}_3 \hat{s}^{*}_{16} \hat{s}^{*}_{36}\right] \; ,
\nonumber \\
\big(M^{}_\nu\big)^{}_{22} \hspace{-0.2cm} & \simeq & \hspace{-0.2cm}
-\left[M^{}_1 \hat{s}^{* 2}_{24} + M^{}_2 \hat{s}^{* 2}_{25} +
M^{}_3 \hat{s}^{* 2}_{26}\right]  \; ,
\nonumber \\
\big(M^{}_\nu\big)^{}_{23} \hspace{-0.2cm} & \simeq & \hspace{-0.2cm}
-\left[M^{}_1 \hat{s}^{*}_{24} \hat{s}^{*}_{34} +
M^{}_2 \hat{s}^{*}_{25} \hat{s}^{*}_{35} +
M^{}_3 \hat{s}^{*}_{26} \hat{s}^{*}_{36}\right] \; , \hspace{0.3cm}
\nonumber \\
\big(M^{}_\nu\big)^{}_{33} \hspace{-0.2cm} & \simeq & \hspace{-0.2cm}
-\left[M^{}_1 \hat{s}^{* 2}_{34} + M^{}_2 \hat{s}^{* 2}_{35} +
M^{}_3 \hat{s}^{* 2}_{36}\right]  \; .
\label{35}
%     (35)
\end{eqnarray}
Let us emphasize that there appear nine CP-violating phases in Eq.~(\ref{35}),
but three of them (or their combinations) are redundant and can always
be removed by rephasing the charged lepton fields in a proper way
%%%%%%%%%%%%%%%%%%%%%%%%%%%%%%%%%%%%%%%%%%%%%%%%%%%%%%%%%%%%%%%%%%%%%%
\footnote{A straightforward way to remove the three redundant phase
parameters of $A$ and $R$ is just to switch off three of the nine phases
in the nine active-sterile flavor mixing matrices $O^{}_{ij}$ (for
$i = 1, 2, 3$ and $j = 4, 5, 6$) in Eq.~(\ref{27}) from the very beginning.
As there are many options in doing so, we do not go into details here.}.
%%%%%%%%%%%%%%%%%%%%%%%%%%%%%%%%%%%%%%%%%%%%%%%%%%%%%%%%%%%%%%%%%%%%%%
A combination of Eqs.~(\ref{34}) and (\ref{35}) allows us to establish
the direct relations between the nine derivational and eighteen original
seesaw parameters. So the former can in principle be determined from the
latter for a given seesaw model (a top-down approach), and the latter
may be partly probed or constrained from the former with the help of some
low-energy neutrino experiments (a bottom-up approach). A careful and
detailed analysis of the parameter space along this line of thought will
be made elsewhere.

\section{Summary}

We have reformulated the canonical seesaw mechanism by considering the
fact that the electroweak gauge symmetry is unbroken at the seesaw
scale characterized by the masses of heavy Majorana neutrinos,
and shown that it can {\it formally} work and allow us to derive an
exact seesaw relation between the active (light) and sterile (heavy)
Majorana neutrinos. In this way we have elucidated the reason why
there is an unavoidable mismatch between the mass eigenstates of heavy
Majorana neutrinos associated with the seesaw and thermal leptogenesis
mechanisms. The smallness of this mismatch has been discussed with the
help of a complete Euler-like parametrization of the flavor structure in
the seesaw framework, and the exact and explicit relations between
the {\it original} and {\it derivational} seesaw parameters have been
established as a by-product.

We hope that this work may help clarify some conceptual ambiguities
associated with the validity of the seesaw mechanism before and after
spontaneous electroweak symmetry breaking, because such ambiguities have
never been taken serious in the literature. It should also be helpful
to clarify the ambiguities associated with the RGE evolution between
the ``virtual" flavor parameters of Majorana neutrinos at the seesaw
scale and those ``real" ones at the electroweak scale, which is crucial
to bridge the gap between a well-motivated UV-complete flavor theory
including the seesaw mechanism and all the possible low-energy flavor
experiments.

\section*{Acknowledgements}

I am greatly indebted to Di Zhang and Shun Zhou for numerous helpful
discussions and comments about this paper, a work dedicated to the 50th
birthday of my home institute, the Institute of High Energy Physics,
which was founded on 1 February 1973. I would also like to thank
Marco Drewes for very useful discussions and comments. My research is
supported in part by the National Natural Science Foundation of
China under grant No. 12075254 and grant No. 11835013.

\newpage

\renewcommand{\theequation}{\thesection.\arabic{equation}}
\section*{Appendix}
\appendix

\setcounter{equation}{0}
\section{The expressions of $A$, $B$, $R$ and $S$}
\label{A}

Given the Euler-like parametrization of the $6\times 6$ unitary flavor mixing
matrix $\mathbb{U}$ decomposed in Eq.~(\ref{27}), the $3\times 3$
active-sterile flavor mixing matrices $A$, $B$, $R$ and $S$ depend on the same
nine rotation angles $\theta^{}_{ij}$ and nine phase angles $\delta^{}_{ij}$
(for $i = 1, 2, 3$ and $j = 4, 5, 6$).
To be explicit~\cite{Xing:2007zj,Xing:2011ur},
\begin{eqnarray}
A \hspace{-0.2cm} & = & \hspace{-0.2cm}
\left( \begin{matrix} c^{}_{14} c^{}_{15} c^{}_{16} & 0 & 0
\cr \vspace{-0.45cm} \cr
\begin{array}{l} -c^{}_{14} c^{}_{15} \hat{s}^{}_{16} \hat{s}^*_{26} -
c^{}_{14} \hat{s}^{}_{15} \hat{s}^*_{25} c^{}_{26} \\
-\hat{s}^{}_{14} \hat{s}^*_{24} c^{}_{25} c^{}_{26} \end{array} &
c^{}_{24} c^{}_{25} c^{}_{26} & 0 \cr \vspace{-0.45cm} \cr
\begin{array}{l} -c^{}_{14} c^{}_{15} \hat{s}^{}_{16} c^{}_{26} \hat{s}^*_{36}
+ c^{}_{14} \hat{s}^{}_{15} \hat{s}^*_{25} \hat{s}^{}_{26} \hat{s}^*_{36} \\
- c^{}_{14} \hat{s}^{}_{15} c^{}_{25} \hat{s}^*_{35} c^{}_{36} +
\hat{s}^{}_{14} \hat{s}^*_{24} c^{}_{25} \hat{s}^{}_{26}
\hat{s}^*_{36} \\
+ \hat{s}^{}_{14} \hat{s}^*_{24} \hat{s}^{}_{25} \hat{s}^*_{35}
c^{}_{36} - \hat{s}^{}_{14} c^{}_{24} \hat{s}^*_{34} c^{}_{35}
c^{}_{36} \end{array} &
\begin{array}{l} -c^{}_{24} c^{}_{25} \hat{s}^{}_{26} \hat{s}^*_{36} -
c^{}_{24} \hat{s}^{}_{25} \hat{s}^*_{35} c^{}_{36} \\
-\hat{s}^{}_{24} \hat{s}^*_{34} c^{}_{35} c^{}_{36} \end{array} &
c^{}_{34} c^{}_{35} c^{}_{36} \cr \end{matrix} \right) \; ,
\nonumber \\
R \hspace{-0.2cm} & = & \hspace{-0.2cm}
\left( \begin{matrix} \hat{s}^*_{14} c^{}_{15} c^{}_{16} &
\hat{s}^*_{15} c^{}_{16} & \hat{s}^*_{16} \cr \vspace{-0.45cm} \cr
\begin{array}{l} -\hat{s}^*_{14} c^{}_{15} \hat{s}^{}_{16} \hat{s}^*_{26} -
\hat{s}^*_{14} \hat{s}^{}_{15} \hat{s}^*_{25} c^{}_{26} \\
+ c^{}_{14} \hat{s}^*_{24} c^{}_{25} c^{}_{26} \end{array} & -
\hat{s}^*_{15} \hat{s}^{}_{16} \hat{s}^*_{26} + c^{}_{15}
\hat{s}^*_{25} c^{}_{26} & c^{}_{16} \hat{s}^*_{26} \cr \vspace{-0.45cm} \cr
\begin{array}{l} -\hat{s}^*_{14} c^{}_{15} \hat{s}^{}_{16} c^{}_{26}
\hat{s}^*_{36} + \hat{s}^*_{14} \hat{s}^{}_{15} \hat{s}^*_{25}
\hat{s}^{}_{26} \hat{s}^*_{36} \\ - \hat{s}^*_{14} \hat{s}^{}_{15}
c^{}_{25} \hat{s}^*_{35} c^{}_{36} - c^{}_{14} \hat{s}^*_{24}
c^{}_{25} \hat{s}^{}_{26}
\hat{s}^*_{36} \\
- c^{}_{14} \hat{s}^*_{24} \hat{s}^{}_{25} \hat{s}^*_{35}
c^{}_{36} + c^{}_{14} c^{}_{24} \hat{s}^*_{34} c^{}_{35} c^{}_{36}
\end{array} &
\begin{array}{l} -\hat{s}^*_{15} \hat{s}^{}_{16} c^{}_{26} \hat{s}^*_{36}
- c^{}_{15} \hat{s}^*_{25} \hat{s}^{}_{26} \hat{s}^*_{36} \\
+c^{}_{15} c^{}_{25} \hat{s}^*_{35} c^{}_{36} \end{array} &
c^{}_{16} c^{}_{26} \hat{s}^*_{36} \cr \end{matrix} \right) \; ; \hspace{0.8cm}
\label{A.1}
%     (A.1)
\end{eqnarray}
and
%which appear in the standard weak charged-current interactions in Eq.~(\ref{1.1}).
%The patterns of $B$ and $S$ are quite analogous to those of $A$ and
%$R$~\cite{Xing:2007zj,Xing:2011ur}:
\begin{eqnarray}
B \hspace{-0.2cm} & = & \hspace{-0.2cm}
\left( \begin{matrix} c^{}_{14} c^{}_{24} c^{}_{34} & 0 & 0
\cr \vspace{-0.45cm} \cr
\begin{array}{l} -c^{}_{14} c^{}_{24} \hat{s}^{*}_{34} \hat{s}^{}_{35} -
c^{}_{14} \hat{s}^{*}_{24} \hat{s}^{}_{25} c^{}_{35} \\
-\hat{s}^{*}_{14} \hat{s}^{}_{15} c^{}_{25} c^{}_{35} \end{array} &
c^{}_{15} c^{}_{25} c^{}_{35} & 0 \cr \vspace{-0.45cm} \cr
\begin{array}{l} -c^{}_{14} c^{}_{24} \hat{s}^{*}_{34} c^{}_{35} \hat{s}^{}_{36}
+ c^{}_{14} \hat{s}^{*}_{24} \hat{s}^{}_{25} \hat{s}^{*}_{35} \hat{s}^{}_{36} \\
- c^{}_{14} \hat{s}^{*}_{24} c^{}_{25} \hat{s}^{}_{26} c^{}_{36} +
\hat{s}^{*}_{14} \hat{s}^{}_{15} c^{}_{25} \hat{s}^{*}_{35}
\hat{s}^{}_{36} \\
+ \hat{s}^{*}_{14} \hat{s}^{}_{15} \hat{s}^{*}_{25} \hat{s}^{}_{26}
c^{}_{36} - \hat{s}^{*}_{14} c^{}_{15} \hat{s}^{}_{16} c^{}_{26}
c^{}_{36} \end{array} &
\begin{array}{l} -c^{}_{15} c^{}_{25} \hat{s}^{*}_{35} \hat{s}^{}_{36} -
c^{}_{15} \hat{s}^{*}_{25} \hat{s}^{}_{26} c^{}_{36} \\
-\hat{s}^{*}_{15} \hat{s}^{}_{16} c^{}_{26} c^{}_{36} \end{array} &
c^{}_{16} c^{}_{26} c^{}_{36} \cr \end{matrix} \right) \; , \hspace{0.8cm}
\nonumber \\
S \hspace{-0.2cm} & = & \hspace{-0.2cm}
\left( \begin{matrix} -\hat{s}^{}_{14} c^{}_{24} c^{}_{34} &
-\hat{s}^{}_{24} c^{}_{34} & -\hat{s}^{}_{34} \cr \vspace{-0.45cm} \cr
\begin{array}{l}
\hat{s}^{}_{14} c^{}_{24} \hat{s}^{*}_{34} \hat{s}^{}_{35}
+ \hat{s}^{}_{14} \hat{s}^{*}_{24} \hat{s}^{}_{25} c^{}_{35} \\
- c^{}_{14} \hat{s}^{}_{15} c^{}_{25} c^{}_{35} \end{array} &
\hat{s}^{}_{24} \hat{s}^{*}_{34} \hat{s}^{}_{35} - c^{}_{24}
\hat{s}^{}_{25} c^{}_{35} & -c^{}_{34} \hat{s}^{}_{35} \cr \vspace{-0.45cm} \cr
\begin{array}{l}
\hat{s}^{}_{14} c^{}_{24} \hat{s}^{*}_{34} c^{}_{35}
\hat{s}^{}_{36} - \hat{s}^{}_{14} \hat{s}^{*}_{24} \hat{s}^{}_{25}
\hat{s}^{*}_{35} \hat{s}^{}_{36} \\ + \hat{s}^{}_{14}
\hat{s}^{*}_{24} c^{}_{25} \hat{s}^{}_{26} c^{}_{36} + c^{}_{14}
\hat{s}^{}_{15} c^{}_{25} \hat{s}^{*}_{35}
\hat{s}^{}_{36} \\
+ c^{}_{14} \hat{s}^{}_{15} \hat{s}^{*}_{25} \hat{s}^{}_{26}
c^{}_{36} - c^{}_{14} c^{}_{15} \hat{s}^{}_{16} c^{}_{26} c^{}_{36}
\end{array} &
\begin{array}{l}
\hat{s}^{}_{24} \hat{s}^{*}_{34} c^{}_{35} \hat{s}^{}_{36}
+ c^{}_{24} \hat{s}^{}_{25} \hat{s}^{*}_{35} \hat{s}^{}_{36} \\
-c^{}_{24} c^{}_{25} \hat{s}^{}_{26} c^{}_{36} \end{array} &
-c^{}_{34} c^{}_{35} \hat{s}^{}_{36} \cr \end{matrix} \right) \; .
\label{A.2}
%     (A.2)
\end{eqnarray}
We see that both $A$ and $B$ are the lower triangular matrices, and the
expression of $B$ can be read off from that of $A^*$ with the
subscript replacements $15 \leftrightarrow 24$, $16 \leftrightarrow 34$ and $26
\leftrightarrow 35$. The expression of $S$ can be similarly obtained
from that of $-R^*$ with the same subscript replacements~\cite{Xing:2020ijf}.
Note, however, that $B$ and $S$ do not affect any physical processes in the
seesaw mechanism.

%\newpage

\end{document}